\documentstyle[12pt]{article}

\font\Bbb=msbm10 scaled 1200

\newcommand{\pl}{\partial_}
\newcommand{\noi}{\noindent}
\newcommand{\ol}{\overline}
\newcommand{\bqq}{\begin{equation}\label}
\newcommand{\eeq}{\end{equation}}
\newcommand{\ptc}{\phi^\tau_C}
\newcommand{\grad}{\,{\rm grad}\,}
\newcommand{\ad}{\,{\rm ad}\,}
\newcommand{\nn}{{\mbox{\Bbb N}}}
\newcommand{\rr}{{\mbox{\Bbb R}}}
\newcommand{\cc}{{\mbox{\Bbb C}}}

\begin{document}
\hsize=15.5cm
\textheight=22cm
\addtolength{\topmargin}{-3.2cm}
\setcounter{page}{0}
\pagestyle{empty}

\rightline{KSU-GRG-97-12}

\begin{center}
{\Large Quantum mechanics in curved space

and quantization of polynomial Hamiltonians}\footnote{
The talk presented at the XVI{\it th} Workshop on Geometric  Methods  in
Physics. June 30 - July 6, 1997, Bialowieza, Poland.}

\bigskip \bigskip

{\bf  Dmitry A. Kalinin}
\medskip

Department of General Relativity \& Gravitation\\
Kazan State University, 18 Kremlyovskaya Ul.\\
Kazan 420008 Russia
\medskip

E-mail: {\tt Dmitry.Kalinin{\it @}\/ksu.ru}
\medskip

July 20, 1997
\vskip3cm

{\bf Abstract}
\medskip
\end{center}

The quantization of a single particle without spin in an appropriate
curved space-time is considered. The Hamilton formalism on reduced
space for a particle in a curved space-time is constructed and the
main aspects of quantization scheme are developed. These
investigations are applied to quantization of the particle Hamiltonian
in an appropriate curved space-time. As an example the energy
eigenvalues in Einstein universe are calculated. In the last sections o
the paper approximation for small values of momenta of the results
previously obtained is considered as well as quantization of polynomial
Hamiltonians of general type is discussed.

\newpage
\pagestyle{plain}

\section{Introduction}

One of the first attempts to construe the quantum mechanics of a
relativistic particle was undertaken by P.~A.~M.~Di\-rac \cite{dirac}
who have considered various forms of relativistic dynamics
corresponding to different types of 3-sur\-fa\-ces in Minkowski
space-time. Dirac considered only the case of the flat space
as far as he supposed that the gravitational effects are insignificant
on the quantum scales. Further development of quantum gravity has
resulted in consideration of quantum systems in which the effects of
general relativity are sufficiently strong.

However, up to the present time doesn't exist the constructive quantum
theory in curved space-time. The difficulties in development of such
theory are connected to the problem of construction the Hamilton
formalism for mechanical systems in curved spa\-ce-ti\-me. They appear
because we have a constrained theory of parametrized type, i.e. each
solution curve of which lies within an phase orbit. Note that the
Einstein's general relativity is parametrized theory too \cite{haj1}).
Therefore, we can hope that the quantization of particle dynamics in
curved space-time could answer questions arising in quantum gravity.

At the present paper we consider quantization of a single particle
without spin in an appropriate curved space-time. There are two
approaches to the quantization of constrained systems (as well as to
the curved space quantization). The first approach corresponds to the
resolution of the constraints at the quantum level and the second
approach is based on the symmetry reduction procedure \cite{mawe1} for
constrained systems (see \cite{DET,MikP} for discussion on the problem
of commutation of reduction and quantization).

Which approach more suitable to the case of quantum particle in curved
space? First approach has the disadvantage that in general case it is
a complicated problem to solve the quantum constraints. However, in
certain cases it is possible to do it by the use of
$3+1$-decomposition \cite{tagirov}.

We propose below a way of quantization which use mostly the second
approach, i.e. solving the constraints at classical level. It has also
one defect that the quantum theory will not be generally covariant
because the Hamiltonian and the Poisson brackets are not covariant
expressions. However, it seems in the case, that using of reduced
phase space method makes physical situation more clear and
calculations more simple.

One more problem we discuss here is the quantization of functions
which are polynomial in momenta (we call them polynomial
Hamiltonians). The example of general relativistic particle (and
quantum gravity) shows the importance of this problem. The approach
which allows to quantize polynomial Hamiltonians on smooth manifolds
is developed in the paper.

The plan of the paper is as follows. In Sec.~\ref{h} the Hamilton
formalism on reduced space for particle in curved space-time is
described. In the next section the main aspects of quantization scheme
(based on geometric quantization approach) are developed. Here we also
describe the existing methods of curved space quantization (for more
complete review see \cite{GotGrTuy,haj1,sniat1,tagirov}) as well as
problems these methods faced to. Sec.~\ref{quant-qs} is devoted to
quantization of general relativistic Hamiltonian in an appropriate
curved space-time. As example the energy eigenvalues of the particle
in Einstein universe are calculated in this section. In
Sec.~\ref{apsec} we consider the approximation of the results
previously obtained for small values of momenta. In the last section
of the paper quantization of polynomial Hamiltonians is discussed.

\section{Hamilton formulation of
\newline
particle dynamics in curved space}\label{h}

We shall not discuss in this paper the aspects of physical definition
of particle in curved space \cite{birdav}. We proceed from the naive
definition of particle as an object which configuration space is a
pseudo Riemannian manifold $({\cal M},g)$ --- the space-time manifold.
Our goal is to develop the Hamilton formulation for
this system.

Let us consider two parts of the Hamilton formulation which are: the
phase space reduction and the reduction at the level of action. First
part results to physical phase space and corresponds to solving of
constraints. The second part should lead to correct general
relativistic Hamiltonian. We consider both these parts in order to
display different features of the system.

To start the discussion we first investigate the phase space
reduction. The phase space of our system is cotangent bundle
$N=T^*{\cal M}$ over space-time manifold ${\cal M}$ with symplectic
2-form $\omega$. Let $\pi: T^* {\cal M}\to {\cal M}$ be the canonical
projection. In a local chart $(\pi^{-1} (U),x^i,p_i)$, $i=0,\ldots,3$,
$U\subseteq {\cal M}$ the 2-form $\omega$ is given by the formula (the
{\it canonical symplectic form})
 \bqq{sform}
\omega=\sum dp_i\land dx^i.
 \eeq
The dynamics of the particle is defined by constraints and
Hamiltonian. The only constraint is the mass-shell condition
$C=1/2(g^*(p,p) - m^2)=0$, $p\in T^*_x {\cal M}$ (here $g^*$ is
induced bilinear form on $T^*_x {\cal M}$). The problem of finding
physical Hamiltonian of the theory will be considered later. Now we
only note that we can choose Hamiltonian in the form $H_{\cal M}=C=1/2
(g^*(p,p)-m^2)$. It is constant on the constraint surface
$\Sigma=\{(x,p)\in N\, |\, C=0\}$ and, hence, phase trajectories in
$N$ preserve $\Sigma$. It means that the considered system is of
parametrized type.

Let $\ad (f)$ be Hamiltonian vector field of the function $f\in
C^\infty (N)$ defined by the condition
 $$
\iota (\ad (f)) \omega = -df, 
 $$
where $\iota$ is the internal product. In local coordinates this
formula reads as
 $$
\ad(f) = \omega^{ij}\partial_j f\partial_i.
 $$
Vector field $\ad (C)$ defines one-parameter transformation
group $\phi^\tau_C$ of sym\-p\-lec\-to\-mor\-p\-hisms in $N$
defined by the equations
 \bqq{Ham-e}
\frac{d\phi^\tau_C x}{d\tau}=\ad (C).
 \eeq
The Lie group $\phi^\tau_C$ acts on symplectic manifold $(N,\omega)$
by symplectomorphisms and preserves the constraint surface $\Sigma$.
Reduction procedure corresponds to the factorization of $\Sigma$
by this action and to the transition to  an orbit space ${\cal O}$
(physical phase space). This space is a symplectic manifold if the
following conditions are obeyed \cite{arnold,mawe1}

1) $\Sigma$ is a smooth manifold,

2) $\phi^\tau_C$ acts on $\Sigma$ without fixed points,

3) the action of group $\phi_C^\tau$ is {\it proper}, i.e
for mapping $(\phi^\tau_C,x)\mapsto (\phi^\tau_C(x),x)$

images of compact sets are compact.

\noi
The first condition is obviously satisfied in the considered case. If
1) and 2) holds then the condition 3) is satisfied also. In order to
prove this fact we note that since the group $\phi_C^\tau$ acts on the
constraint surface $\Sigma$ without fixed points then any orbit
$\gamma$ of this action can be identified with the group $\ptc$. By
choosing a suitable reparametrization $\tau'=\tau' (\tau)$ one can
derive that the action of group $\ptc$ on $\gamma$ corresponds to the
left-shift action of the group on itself. Because left-shift action is
proper, the condition 3) is satisfied.

Let us pass to the second condition. It is satisfied if and only if the
vector field $\ad (C)$ is nonvanishing on $\Sigma$. In a local chart
this means
 \bqq{neq}
g^{\alpha\beta} p_\beta\ne 0 \quad \mbox{or} \quad
\pl\sigma g^{\alpha\beta} p_\alpha p_\beta \ne 0.
 \eeq
If $m\ne 0$ (as it is in the case of massive particle) then $p\ne 0$
and (\ref{neq}) is satisfied. If $m=0$ (the case of
massless particle) then in order to satisfy the second condition one
should consider the space $\Sigma\backslash \Sigma_0$ where
$\Sigma_0=\{(x,0)\in T^* {\cal M}\}$, $\Sigma_0\subset \Sigma$. So we
prove that conditions 1) -- 3) are satisfied and Hamilton formulation
can be obtained by reduction of the initial system.

For physical applications the concrete realization of the reduction
procedure, i.e. the orbit space construction have
exceptional significance. For this purpose we shall use {\it
transversal surface} \cite{arnold,haj2,haj1} $N_{\rm tr}$ which is in
our case the cotangent bundle $T^* M_0$ for some spacelike
3-sur\-fa\-ce $M_0\subset {\cal M}$. We shall call $M_0$ {\it instant
{\rm 3}-sur\-fa\-ce}. Transversal surface is realization of the
orbit space, i.e. it is transversal to all orbits and intersects each
orbit only in one point. Moreover, $N_{\rm tr}$ is a symplectic
manifold with the symplectic form projected from $N$.

A smooth function $T:{\cal M}\to \rr$ increasing along any
non-spacelike curve is called {\it global time function}
\cite{beemehr}. A Lorentzian manifold $({\cal M},g)$ admits the global
time function if and only if $({\cal M},g)$ is {\it stably
casual}\footnote{It means that there are no closed timelike or null
curves in any Lorentzian metric which is sufficiently near to the
metric $g$ in the sense of some $C^0$ open topology on the space $Lor
({\cal M})$ of all $C^2$ Lorentz metric on the space-time manifold
${\cal M}$ \cite{haw1}.}. In general case there are no natural way of
choosing global time function on a stably casual space-time. Let
$\grad (f)$ be the gradient flow of a smooth function $f: {\cal M}\to
\rr$, i.e. $g(\grad (f),X)=df (X)$ for any vector field $X$ on ${\cal M}$.
If $\grad (f)$ is time-like, then $f$ increases or decreases along any
non-spacelike curve and $f$ or $-f$ can be taken as global time
function.

Let a space-time $({\cal M},g)$ be stably casual with the global time
function $T$. Then we can take the instant 3-surface in the form $M_0
= \{x\in {\cal M}\, | \, T(x)=T_0=\mbox{const} \}$. Let $x^i=x^i
(v^\mu)$ are the local equations of embedding for $M_0$, then we get
the reduced phase space $T^* M_0$ with the canonical symplectic form
 $$
\omega=\sum dp_\mu\land dv^\mu, \qquad \mu=1,2,3.
 $$

The reduction of phase space doesn't give us the Hamiltonian of the
theory. In order to solve this problem we shall consider the action of
the system. In the case of scalar particle in curved space ${\cal M}$
it can be given by the following formula
 \bqq{action1}
S[x]= m\int dt \sqrt{g(\dot x,\dot x)}.
 \eeq
This action has to be supplemented by the mass-shell condition
 \bqq{H-1}
g(\dot x, \dot x)=m^2
 \eeq
which is the constraint of the theory.
Let us rewrite the action in the form
 \bqq{action2}
S[x,e] = 1/2 \int d\tau ( \frac{g(\dot x,\dot x)}{e} + e m^2).
 \eeq
It is easy to prove that these two forms of action define equivalent
forms of dynamics. Now we can obtain from (\ref{action2}) the primary
constraint $p_e=0$ and the canonical Hamiltonian
 \bqq{canH}
H_{\rm can}=eH_{\cal M}, \qquad H_{\cal M}=\frac{1}{2}(g^*(p,p)-m^2).
 \eeq
The Poisson brackets of the primary constraint and the Hamiltonian
gives the secondary constraint
 \bqq{sec-cons}
\{p_{e},H_{\rm can}\}=H_{\cal M}=0.
 \eeq
Now we can rewrite the initial action (\ref{action2}) in the
Hamiltonian form
 \bqq{inteta}
S = \int_\gamma \Theta
 \eeq
where $\Theta=\theta - e H_{\cal M} dt$ is {\it Poincar\`e-Cartan
integral invariant} \cite{arnold} and $\theta=p_0 dT +p_\mu dv^\mu$ is
{\it action} 1-{\it form} on the phase space $N$ defined by the
condition $\omega=d\theta$ up to addition of an exact 1-form $df$. The
integral in (\ref{inteta}) should be taken along the graph $\gamma$ of
the system with Hamiltonian $H_{\cal M}$ in evolution space $N\times
\rr$.

Using (\ref{sec-cons}), we get $S=\int_{\tilde\gamma}\theta|_\Sigma$
where $\tilde\gamma$ is the projection of $\gamma$ on the reduced phse
space $N_{tr}$

Let us choose new coordinate system $(x^A,C,\varphi)$, $A=1,\ldots,6$
so that
           $$
q^i=q^i (x^A,C,\varphi), \qquad
p^i=p^i (x^A,C,\varphi).
           $$
Then we can write $\theta$ in the form
           $$
\theta=p_i \frac{\partial q^i}{\partial x^A} dx^A
+p_i \frac{\partial q^i}{\partial C}dC+
+p_i \frac{\partial q^i}{\partial \varphi}d\varphi,
           $$
hence,
           $$
\theta|_\Sigma = (\zeta_A \frac{d x^A}{d \varphi}-H_\varphi) d\varphi
           $$
where
           \bqq{fA}
\zeta_A=p_i \frac{\partial q^i}{\partial x^A}, \qquad
H_\varphi= -p_i \frac{\partial q^i}{\partial \varphi}.
           \eeq
Now we can write down the action in new coordinates
           $$
S = \int
\zeta_A \frac{d x^A}{d\varphi} - H_\varphi)d\varphi.
           $$
We find the Hamiltonian form for action of the system with the
Hamiltonian (\ref{fA}). The phase space of this system is $N_{tr}$
whose symplectic form (locally) is
           $$
\omega= \frac{\partial\zeta_B}{\partial x^A} dx^A\land dx^B.
           $$
Here the function $H(\varphi,x^1,\ldots,x^6)$ is defined by
 $$
C(-F,H,x^1,\ldots,x^6)=0.
 $$

As an example we consider the instant form of particle dynamics in
curved space-time ${\cal M}$ (which is rather similar to instant
dynamics in Minkowski space \cite{dirac}). Let ${\cal M}$ be a stably
casual Lorentzian manifold with global time function $T$ and
instant 3-sur\-fa\-ce $M_0$. Let $v^\mu$ are the local
coordinates on $M_0$. If we take $(T,v^\mu)$ as coordinate system in
${\cal M}$ then $g_{0\mu}=0$, $g_{00}>0$ and the functions
$\gamma_{\alpha\beta} = - g_{\alpha\beta} / g_{00}$,
$g^{\alpha\mu}g_{\mu\beta} = \delta^\alpha_\beta$ are the components
of Riemannian metric on $M_0$. Taking $\varphi=p_0$ we find from
(\ref{fA})
 $$
H = \pm \sqrt{\gamma^{\mu\nu}p_\mu p_\nu + g_{00} \: m^2}.
 $$
As the result, we get the following reduced action form
 $$
S_{\rm red} = \int dT
(p \frac{dv}{dT}-H), \quad (v,p)\in N_{\rm tr}=T^*M_0.
 $$
From here it follows that $H$ is the Hamiltonian of the mechanical
system with phase space $N_{\rm tr}=T^*M_0$. This system is equivalent
to the initial sytem with Hamiltonian $H_{\cal M}=1/2 (g^*(p,p)-m^2)$
and the constraint $C=H_{\cal M}=0$.

It is possible to construct another forms of dynamics by choosing
different types of 3-surfaces (timelike or null) in the reduction
procedure. This will result to another choice of the function
$\varphi$. These forms of dynamics could be useful for systems which
don't admit the description in terms of Hamilton mechanics on {\it
instant} 3-sur\-fa\-ce but {\it do admit} Hamilton formulation achieved
by the use of reduction procedure (in fact for such systems should be
chosen another representation for orbit spaces). In the following we
shall restrict ourselves only to the consideration of the instant form
of dynamics.

\section{Geometric quantization}\label{gq}

In this section we describe the geometric quantization approach. Later
it will be used for quantization of previously considered system. Let
us start from formulation of quantization axioms
\cite{hurt,kir2,sniat1,Woodhouse}.

{\it Quantization} is the linear map ${\cal Q}:f \mapsto \hat f$ of
Poisson (sub)algebra $C^{\infty} (N)$ into the set of operators in
some (pre)Hilbert space ${\cal H}$ possessing the following properties:

(Q1) $\hat 1=1$;

(Q2) $\widehat {\{ f,g\}}_h = \frac{i}{h}(\hat f \hat g - \hat g \hat
f)$;

(Q3) $\hat {\ol f} = (\hat f)^{*}$;

(Q4) for a complete set of functions
$f_1,\ldots,f_n$ the operators ${\hat  f}_1 ,\ldots,{\hat f}_n$

also form a complete set.

\noindent Here the bar is the complex conjugation, the star denotes
the conjugation of operator and $h= 2\pi\hbar$ is the Planck constant.

{\it Geometric quantization approach} was developed
\cite{kir2,hurt,sniat1} to help solve some difficulties of canonical
quantization scheme. These difficulties are especially essential in
the case of the phase spaces with non-tri\-vi\-al topology
\cite{GotGrTuy,hurt}.

Let $(N,\omega)$ be a $n$-dimensional symplectic manifold. Geometric
quantization procedure consists of the following main parts.

a) {\it Prequantization line bundle} ${\cal L}$ that is a Hermitian
line bundle over $N$ with connection $D$ and
$D$-in\-va\-ri\-ant\footnote{Recall that $D$-invariance means that for
each pair of sections $\lambda$ and $\mu$ of ${\cal L}$ and each real
vector field $X$ on $M$ holds $X<\lambda ,\mu >=<D_X \lambda
,\mu>+<\lambda,D_X \mu>$.} Hermitian structure $<,>$. The curvature
form $\Xi$ of the prequantization bundle ${\cal L}$ have to coincide with
the form $h^{-1}\omega$.

b) {\it Polarization} $F$ that is an involutive Lagrange distribution
in $TN\otimes_{\rr}{\cc}$.

c) {\it Metaplectic structure} which consists of the {\it bundle of
metalinear frames in} $TN$ and the {\it bundle ${\cal L}\otimes
\sqrt{\land^n F}$ of ${\cal L}$-va\-lu\-ed half-forms normal to the
polarization $F$} (see, for example, \cite{sniat1}).

If these three structures are defined on ($M,\omega$), then the
Hilbert space ${\cal H}$ of the system consists of such sections $\mu$
of the bundle ${\cal L}\otimes \sqrt{\land^n F}$ which are covariantly
constant with respect to the polarization $F$.
The {\it Souriau-Kostant prequantization formula}
    \bqq{SKQ}
{\cal P} (f) = f-i\hbar D_{\ad (f)}.
    \eeq
provides the set of the operators ${\cal P}(f)$ satisfying the first
three quantization axioms. In order to satisfy the axiom (Q4) one
should introduce the polarizations and use the {\it
Blat\-t\-ner-Cos\-tant-Ster\-n\-berg (BKS) kernel}. The BKS kernel is
a bilinear mapping $K: {\cal H}_{F_1} \times {\cal H}_{F_2} \to \rr$
which connects Hilbert spaces for two different polarization $F_1$ and
$F_2$.

It is possible to offer two approaches for the quantization of a
mechanical system. The first approach corresponds to the resolution of
the constraints at the quantum level and the second approach is based
on the Hamilton formalism for constrained systems reviewed in Sec.~2.
What this two approaches give us in application to general
relativistic particle case?

If we use the quantum constraints then we have to start from the phase
space $(T{\cal M}, \omega$) where $\omega$ is given by
Eq.~(\ref{sform}). Then the operator ${\cal Q}(H)$ corresponding to
the Hamiltonian (\ref{canH}) allocates admissible (physical) states in
the Hilbert space ${\cal H}$. If we write the classical constraint in
the form $C^2=0$, then the quantum constraint is ${\cal Q}(C^2)=0$.
Geometric quantization gives the following expression for this operator
\cite{dewitt,sniat1}\footnote{The similar expression for quantization
of $C^2$ was obtained using another approach in \cite{DET}.}
     \bqq{l}
{\cal Q}(C^2)\: \psi\: \lambda_0 \otimes \mu_0 = (-\hbar^2 (g^{ij}
\nabla_i \nabla_j - \frac{1}{6} R) +m^2)\: \psi\: \lambda_0
\otimes \mu_0
     \eeq
where $R$ is the Ricci scalar, $\psi\in C^\infty (M_0)$ and $\lambda_0
\otimes \mu_0$ is a special nonvanishing section of the bundle ${\cal
L}\otimes \sqrt{\land^n F}$ (see the next section). The last formula
means that physical states have to obey the
con\-for\-mal\-ly-in\-va\-ri\-ant Klein-Gor\-don-Fock equation
\cite{chertag}. Solutions of this equation form {\it physical Hilbert
space} of our system. In general case it is a complicated problem to
solve this quantum constraint. However, in certain cases it is
possible to solve this problem by the use of $3+1$ decomposition
\cite{tagirov}.

At the sane time, solving the constraints at classical level has also
the disadvantage that the quantum theory will not be generally
covariant in this case because the Hamiltonian and the Poisson
brackets are not covariant expressions (such situation reminds the
gauge fixing in quantum gauge theories). However, using of reduced
phase space makes (in certain cases) physical situation more clear and
calculations more simple. It allows to solve problems of quantum
mechanics in curved space which would be rather complicated if we
would use the first approach.

One more way of solving the constraint at classical level could be
proposed. It is based on the fact that the {\it constraint surface}
$\Sigma$ is a {\it Poisson manifold} and it is possible to use the
method of geometric quantization of Poisson manifolds proposed by
I.~Vaisman \cite{vaisman} (see also \cite{rovelli}).


\section{Quantization in curved space-time}\label{quant-qs}

We consider here (see also \cite{kalinin}) geometric quantization of
the system investigated in Sec.~\ref{h} at classical level. Let
$({\cal M},g)$ be a space-time and $M_0$ is the instant surface in
${\cal M}$ (we note here that $M_0$ is orientable). Let $\gamma$ be
the Riemannian metric on $M_0$ projected from $({\cal M},g)$.

The metric $\gamma$ can be used to define a global section of the
bundle $\sqrt{\land^n F}$. We give here short review of this
construction, see \cite{sniat1} for complete reference.

First, one can note that there exists isomorphism between the sections
$\mu$ of ${\bf L}\equiv \sqrt{\land^n F}$ and the set of complex
valued functions $\nu^{\#}$ on the bundle ${\bf L}^*\equiv {\bf
L}\backslash \{ 0 \}$ possessing the property
   $$
\mu^{\#} (cz)=c^{-1} \mu^{\#} (z), \quad c\in \cc.
   $$
This isomorphism is given by the formula
   $$
\mu (\pi z)= \mu^{\#} (z)z
   $$
where $\pi$ is the projection of the bundle ${\bf L}$. If $\{\eta_i\}$
is a metalinear frame \cite{sniat1}, corresponding to the polarization
$F$, then we can define a section $\tilde\mu$ of ${\bf L}$ by the
equation $\tilde\mu^{\#}\circ \eta=1$. Let $U\subset N$ be an open
domain. We introduce a local section $\mu_0 |_U$ of ${\bf L}$ over $U$
by
   $$
\mu_0 |_U= \pm |(\det\gamma)\circ \pi|^{1/4}\tilde\mu.
   $$
Using well-known transformation properties of the metric tensor
$\gamma$ and the half-form $\tilde\mu$ it can be shown that the
sections $\mu_0 |_U$ define {\it global section} $\mu_0$ of ${\bf L}$
which doesn't depend on the choice of particular coordinates.

Let us introduce in $T^* M_0$ {\it vertical polarization} $F$ spanned
by the $\partial /\partial v^\alpha$ where $v^\alpha$, $\alpha=1,2,3$
are local coordinates in $M_0$. The Hilbert space ${\cal H}$
corresponding to this polarization consists of the sections of the
form $\psi (v^\alpha) \lambda\otimes\mu_0$ \cite{sniat1}. The
dynamical variables of physical interest are the canonical coordinates
$v^\alpha$, the corresponding momenta $p^\alpha$ and the Hamiltonian
$H$. Coordinates and momenta preserve the polarization and can be
easily quantized by the use of Sou\-ri\-au-Kos\-tant prequantization
formula (\ref{SKQ}) which yields
   $$
{\cal Q}(v^\alpha)=v^\alpha \quad \mbox{  and } \quad
{\cal Q} (p^\alpha)=\frac{i}{\hbar} \frac{\partial}{\partial v^\alpha}.
   $$

However, the Hamiltonian $H$ doesn't preserve the polarization.
Hence, it is necessary to use BKS kernel to quantize $H$.

As the first example let us consider
the problem of finding energy eigenvalues of scalar particle in {\it
Einstein universe}. In this case it is convenient to quantize the
square $H^2$ of the Hamiltonian but not the Hamiltonian itself. The
spa\-ce-ti\-me ${\cal M}$ in this case is isomorphic to $S^3\times
\rr$ with the metric
 $$
 ds^2=dt^2-a^2 d\Omega^2
 $$
where $a\in \rr$ and $d\Omega^2=\sigma_{\mu\nu}dx^\mu dx^\nu$ is the
metric on 3-sphe\-re $S^3_1$ of unit radius. Hamiltonian of the system
can be taken in the form
 \bqq{EinH}
 H=(a^{-2}\sigma^{\mu\nu} p_\mu p_\nu+m^2)^{1/2}.
 \eeq
Similarly with (\ref{l}) we find
 \bqq{l-E}
 {\cal Q}(H^2) \psi \ \lambda_0 \otimes \mu_0 =
 (-\hbar^2 (a^{-2}\sigma^{\mu\nu} \nabla_\mu \nabla_\nu
 -\frac{1}{6} R) +m^2)\psi \ \lambda_0 \otimes \mu_0
 \eeq
where covariant derivatives and Ricci curvature should be calculated
for 3-met\-ric $d\Omega^2$. The eigenvalues of the operator ${\cal
Q}(H^2)$ coincides with squares $\lambda^2$ of energy eigenvalues
$\lambda$. We shall now calculate $\lambda$'s using the fact that
eigenvalues $\rho_k$ of the Laplacian $\bigtriangleup$ on 3-sphe\-re
$S^3_a$ of the radius $a$ are \cite{bgm}
 $$
\rho_k=-\frac{k(k+2)}{a^2}, \qquad k\in \{0\}\cup \nn.
 $$
The Ricci curvature of $S^3_a$ is equal to $R=6 a^{-2}$. From here it
follows that eigenvalues of the operator ${\cal Q}(H^2)$ (which
coincide with the stationary energy levels of quantum scalar particle
in Einstein universe) are
 \bqq{eig}
\lambda_l=(\hbar^2\frac{l^2}{a^2}+m^2)^{1/2},
\qquad l\in\nn.
 \eeq

Note, that if we consider the Hamiltonian $H_{\rm FRW}$ of {\it
massless} scalar particle in closed
Fri\-ed\-mann-Ro\-bert\-son-Wal\-ker space-time with the metric
$ds^2=dt^2-a(t)^2 d\Omega^2$ then because of conformal invariance of
quantum Hamiltonian it is easy to see that the energy eigenvalues in this
case are also given by (\ref{eig}) (with $m=0$ in this case).

Now we quantize the Hamiltonian $H$ in more general case. To simplify
the calculations in the following we shall consider only the case when
$g_{00}=1$. Let $N = T^*M_0$ be reduced phase space of our system. The
most general quantization formula for an appropriate function $f\in
C^\infty (N)$ in geometric quantization method can be written in the
following form \cite{sniat1}
 \bqq{q}
{\cal Q} (f) \sigma = i\hbar \frac{d}{dt}
(\tilde\phi^t_f\sigma)|_{t=0},\qquad \sigma\in {\cal H}.
 \eeq
Here $\tilde\phi_f^t$ is one-pa\-ra\-me\-ter group of transformation
of Hilbert space ${\cal H}$ induced by the function $f$ via Poisson
brackets. If $\lambda_0$ is nonvanishing local section of
prequantization bundle ${\cal L}$, then $\sigma=\psi\:
\lambda_0\otimes\mu_0$. Using BKS kernel method it is possible to
rewrite (\ref{q}) in the form
 \bqq{q2}
{\cal Q} (f) [\psi\: \lambda_0\otimes\mu_0]
= i\hbar \frac{d}{dt} \psi_t|_{t=0}\lambda_0 \otimes\nu_{\tilde\xi},
\eeq
$$
\psi_t (v^\alpha)=(i\hbar)^{-3/2}\int [\det \omega (\ad
(v^\mu), \phi^t_f \ad (v^\nu))]^{1/2}
 $$
 \bqq{p-t}
\exp [i/\hbar \int^t_0 (\theta (\ad (f))-f)\circ \phi^{-s}_f ds]
\psi (v^\alpha\circ \phi^{-t}_f) d^n p.
 \eeq
where $\omega=d\theta$ and $\theta$ is the action 1-form, chosen as
folows: $\theta=p_\alpha dv^\alpha$. For $f=H$ we have
 \bqq{lagr}
 L=\theta (\ad (H))-H=\frac{-m^2}{H}.
 \eeq
Note, that for $m=0$ the formula (\ref{lagr}) indicates that the
system has zero Lagrangian.

The orbits of the one-pa\-ra\-me\-ter group $\phi^t_{H}$ project onto
geodesics in $M_0$
 $$
 \frac{d^2}{dt^2} (v^\alpha\circ \phi^t_{H})=
 -\Gamma^\alpha_{\mu\nu}\circ \phi^t_{H}
 \frac{d}{dt} (v^\mu\circ \phi^t_{H})
 \frac{d}{dt} (v^\nu\circ \phi^t_{H})
 $$
where $\Gamma^\alpha_{\mu\nu}$ are Christoffel symbols of the metric
$\gamma$. The integrand in (\ref{p-t}) can be simplified if the
coordinates $v^\alpha$ are normal in a point $y_0\in N_{\rm tr}$. In
this case the functions $v^\alpha$ depend linearly on parameter $t$
along the geodesics originating in $y_0$: $v^\alpha\circ\phi^t_H=t
v^\alpha\circ \phi^1_H$ and $\gamma_{\alpha\beta} =
\delta^\alpha_\beta$. In this coordinates
$\Gamma^{\alpha}_{\mu\nu}|_{y_0} = 0$,
$\pl\alpha\gamma_{\mu\nu}|_{y_0} = 0$.

We approximate the integrand in (\ref{p-t}) so that the integration
will give results accurate to first order in $t$. For this, it
suffices to approximate the integrand up to first order in $t$ and
order 2  in $tp^\alpha$. Equation $\frac{d}{dt} (v^\alpha\circ
\phi^t_H) =\gamma^{\alpha\beta} p_\beta H^{-1}$ written in the
normal coordinates yields
 \bqq{1}
v^\alpha\circ \phi^{-t}_{H}=-\frac{t\gamma^{\alpha\beta}p_\beta}{H}
+ \mbox{higher order terms}.
 \eeq
From here we find \cite{sniat1}, p.~132
 $$
\phi^{t}_H\ad (v^\alpha)=\ad (v^\alpha\circ \phi^{-t}_H),
 $$
hence,
 $$
\omega (\ad(v^\alpha), \phi^t_H \ad (v^\beta))= [\ad (v^\alpha)
(v^\beta-\frac{t\gamma^{\beta\mu}p_\mu}{H})] + \mbox{higher order terms}=
 $$
 $$
t\delta^\alpha_\beta H^{-3}+ tp_\alpha p_\beta H^{-3}+ \mbox{higher
order terms}
 $$
and
 \bqq{detom}
(\det(\omega (\ad(v^\beta), \phi^t_H\ad (v^\gamma)))^{1/2}=
t^{3/2} m (\sum_\sigma (p_\sigma)^2 +m^2)^{-5/4} +
\mbox{higher order terms}.
\eeq
In the following considerations we restrict ourselves only to the case
$m\neq 0$ so we can omit next $t$-term in this expression.

Using (\ref{lagr}), (\ref{1}) and (\ref{detom}) we can rewrite
(\ref{p-t}) in the following form
 \bqq{2}
\psi_t (0)=(i\hbar)^{-3/2}\int t^{3/2} m(\sum_\sigma (p_\sigma)^2 +
m^2)^{-5/4} \psi (\frac{-tp_\alpha}{\sqrt{\sum (p_\sigma)^2 + m^2}})
d^3 p.
 \eeq
This formula defines the quantum Hamiltonian of scalar particle in
curved space-ti\-me.

The Hamiltonian has the complicated form and it is hard to calculate
it directly. However, it is possible consider some approximation of
the Hamiltonian in order to simplify the calculations. In the next
section we investigate a lower order (with respect to momenta)
expansion of the Hamiltonian. Using geometric quantization approach it
is in principle possible to calculate the quantum operator
corresponding to each term of the lower order expansion. We calculate
such the operators for the first three terms.


\section{Approximation for small values of momenta}\label{apsec}

In this section an approximation of the formula (\ref{2}) for small
values of momenta will be given. By small momenta we mean the points
of the reduced phase space $T^*M_0$ obeying
$\gamma^{\alpha\beta}p_\alpha p_\beta << m^2$. In this approximation
the Hamiltonian $H$ takes the form
   \bqq{appr}
H=H_1+H_2+H_3+\mbox{higher order terms}.
   \eeq
where
   $$
H_1=m\sqrt{g_{00}},\qquad
H_2=\sqrt{g_{00}}\;\frac{g^{\alpha\beta}p_\alpha p_\beta}{2m}, \qquad
H_3=m\sqrt{g_{00}}\;\frac{(g^{\alpha\beta}p_\alpha p_\beta)^2}{4 m^3}.
   $$
Let us quantize the function $H$ given by the formula (\ref{appr}).
For simplicity we consider only the case $g_{00}=1$. Quantizing
the first two terms in this expression we get
   \bqq{ftt}
m -\frac{\hbar^2}{2m}g^{\alpha\beta}
\nabla_\alpha \nabla_\beta +\frac{\hbar^2}{12m} R
    \eeq
where $\nabla_\alpha$ is the covariant derivative with respect to
3-metric $g_{\alpha\beta}$ and $R$ is the scalar curvature.

Let us consider the third term in (\ref{appr}). In normal coordinates
we have $g_{\alpha\beta}(0)=\delta^\alpha_\beta$. Since $\pl\gamma
g_{\alpha\beta}(0)=0$ in normal coordinates we do not take in
account higher order terms (which are always proportional to first
derivatives), then, according to (\ref{q2}), the BKS-kernel
quantization of $(g^{\alpha\beta}p_\alpha p_\beta)^2$ coincides with
the quantization of the function $\sum (p_\alpha)^4$.

Let $r$ be a real polynomial. It is easy to demonstrate
\cite{GotGrTuy}, that the following {\it higher degree Von Neumann
rule} follows from the general quantization axioms (see Sec.~3)
     $$
{\cal Q}(r(p))=r({\cal Q}(p)).
     $$
From here we have
     \bqq{h3}
{\cal Q}\:(H_3)\:\psi (v)=\frac{\hbar^4}{4m^2} \sum_{\alpha}
\frac{\partial^4 \psi (v)}{\partial v_\alpha^4} (0).
     \eeq
Taking in account the equation \cite{sniat1}
     \bqq{detg}
\psi (v^\alpha (y))=\psi(y) |\det g(y)|^{1/4}.
     \eeq
we can find the expression for $({\cal Q}H_3)\psi$ in an appropriate
coordinates.

In order to do it we first list all possible independent invariant
terms which could appear in this expression:
     $$
I_1=R^2\psi,\quad I_2=\bigtriangleup R\psi, \quad
I_3= R^{\alpha\beta\mu\nu} R_{\alpha\beta\mu\nu}\psi,
     $$
     \bqq{terms}
I_4= R^{\alpha\beta} R_{\alpha\beta},\quad I_5=
R^\alpha\psi_{,\alpha},\quad I_6 = R^{\alpha\beta}
\psi_{,\alpha\beta},\quad
     \eeq
     $$
I_7= R\bigtriangleup \psi,\quad I_8=g^{\alpha\beta}
g^{\mu\nu} \psi_{,\alpha\beta\mu\nu}.
     $$
Here the comma denotes the covariant derivative $\nabla_\alpha$,
$R_{\alpha\beta\mu\nu}$, $R_{\alpha\beta}$ and $R$ are Riemann, Ricci
and scalar curvatures and
$\bigtriangleup= g^{\alpha\beta} \nabla_\alpha
\nabla_\beta$. Using (\ref{detg}), in normal coordinates we find from
(\ref{h3})
     $$
g^{\alpha\beta} g^{\mu\nu} \pl{\alpha\beta\mu\nu}(|g|^{1/4}\psi)=
g^{\alpha\beta} g^{\mu\nu} (|g|^{-3/4} \pl{\beta\mu}|g|
\pl{\alpha\lambda}\psi +|g|^{-3/4}\pl{\beta\lambda\mu}|g|\pl\alpha\psi
+
     $$
     \bqq{main}
\frac{1}{2}|g|^{-3/4}\pl{\lambda\mu}|g|\pl{\alpha\beta}\psi +
(-\frac{3}{8}|g|^{-7/4}\pl{\alpha\lambda}|g|\pl{\beta\mu}\psi -
\frac{3}{16}|g|^{-7/4}\pl{\alpha\beta}|g|\pl{\lambda\mu}\psi
     \eeq
     $$
+\frac{1}{4}|g|^{-3/4}\pl{\alpha\beta\lambda\mu}|g|)\psi
+|g|^{-1/4}\pl{\alpha\beta\lambda\mu}\psi)
     $$
where $|g|=\det (g_{\alpha\beta})$.

Using the following expansion for the components of the metric $g$ in
normal coordinates \cite{petrov}
     $$
g_{\mu\nu}=g_{\mu\nu}(0)+ \frac{1}{3}R_{\mu\alpha\nu\beta}(0)v^\alpha
v^\beta - \frac{1}{6}R_{\mu\alpha\nu\beta,\gamma}(0)v^\alpha
v^\beta v^\gamma +
     $$
     $$
(\frac{1}{20}R_{\mu\alpha\nu\beta,\gamma\lambda}(0)+
\frac{2}{45}R_{\alpha\mu\beta\sigma} R^\sigma_{\gamma\nu\lambda}(0))
v^\alpha v^\beta v^\gamma v^\lambda + \mbox{higher order terms},
     $$
Using this formula, after complicated but straightforward
calculations, we get
     $$
{\cal Q}\: (H_3)\: \psi\: \lambda_0\otimes\mu_0=
|g|^{1/4}(-\frac{1}{12}I_1+ \frac{1}{5}I_2+ \frac{3}{10}I_3+
\frac{1}{30}I_4- \frac{1}{3}I_5+ \frac{1}{3}I_7+
I_8)\lambda_0\otimes\mu_0.
     $$

Using the proposed approach it is in principle possible to calculate
quantum operators corresponding to all terms in the lower order
expansion (\ref{appr}) of the curved space particle Hamiltonian. In
order to do it, for example, for the term $(g^{\alpha\beta}p_\alpha
p_\beta)^3$ one should calculate all possible invariants of 6{\it th}
order (the analog of (\ref{terms})) in normal coordinates and, then
rewrite the expression
     $$
g^{\alpha\beta} g^{\lambda\mu} g^{\nu\rho}
\partial_{\alpha\beta\lambda\mu\nu\rho} (|g|^{1/4}\psi)
     $$
in a covariant form.


\section{Quantization of polynomial Hamiltonians}

Let $M$ be a smooth manifold and $f\in C^\infty (T^* M)$ is an
observable for a dynamical system with phase space $T^* M$. In some
cases it can be usefull to expand $f$ in neighborhood of the points
$p=0$ (i.e. for small values of momenta
     \bqq{appr1}
f=f_{(0)}(q)+f_{(1)}(q)_{\alpha}p^\alpha+f_{(2)}(q)_{\alpha\beta}
p^\alpha p^\beta + f_{(3)}(q)_{\alpha\beta\gamma}p^\alpha p^\beta
p^\gamma + \ldots .
     \eeq

In order to construct Hilbert space operator for such kind of the
functions we propose the following procedure of quantization of the
functions polynomial in momenta which we will call {\it "polynomial
Hamiltonians"}.

First component of this procedure is the global section
$\lambda_0\otimes\mu_0$ of the bundle ${\cal L}\otimes\sqrt{\land^n
F}$ of ${\cal L}$-va\-lu\-ed half-forms normal to (vertical)
polarization. As it was said in Sec.~\ref{quant-qs}, such section can
be constructed using a non-degenerate 2-form field on the
configuration space.

In fact we can choose {\it any} bilinear form $g$ to construct the
section $\lambda_0\otimes\mu_0$. But in applications we usually
interest in {\it natural choice} of $g$.

The problem simplifies if the form $f_{(2)}(q)_{\alpha\beta}$ in
Eq.~(\ref{appr1}) is non-degenerate. In this case the section
$\lambda_0\otimes\mu_0$ can be defined using this form. However, if
$f_{(2)}$ is degenerate one should choose $\lambda_0\otimes\mu_0$
using physical considerations.

When global section of ${\cal L}\otimes\sqrt{\land^n F}$ is chosen one
has to develop some approach for quantization of the terms in
(\ref{appr1}) for any order. The first two terms leave the vertical
polarization invariant and so we can write down the corresponding
operators
     $$
f = f_{(0)}+f_{(1)}(q)_\alpha
p^\alpha,
     $$
     $$
{\cal Q}(f)\psi\: \lambda_0\otimes\mu_0= (-i\hbar \ad
(f)\psi+(f_{(0)}(q)-\frac{i\hbar}{2} \frac{\partial
f_{(1)}(q)_\alpha}{\partial q^\alpha})\psi)\lambda_0\otimes\mu_0.
     $$

Let us consider quantization the terms of order 2 and higher in
momenta. As an example we take the function
     $$
f_n = f_{(n)}^{\alpha_1 \alpha_2 \ldots \alpha_n}
p_{\alpha_1} p_{\alpha_2} \ldots p_{\alpha_n}
     $$
of order $n$.

Let us introduce in a neighborhood of a point $x\in M$ coordinates
such that first derivative of $f_{(n)}^{\alpha_1 \alpha_2 \ldots
\alpha_n}$ in $x$ are equal to zero. We call such coordinates {\it
normal with respect to} $f_n$. Then it is easy to see that the
BKS-kernel quantization of $f_n$
coincides with the quantization of the function
    $$
f_{(n)}^{\alpha_1 \alpha_2 \ldots \alpha_n}|_x
p_{\alpha_1} p_{\alpha_2} \ldots p_{\alpha_n}.
    $$
From here we find
    $$
{\cal Q} (f_n) \psi (v)= f_{(n)}^{\alpha_1 \alpha_2 \ldots \alpha_n}|_x
\frac{\partial^n \psi (v)}{\partial{\alpha_1}\partial{\alpha_2} \ldots
\partial{\alpha_n}}(0).
    $$
Using (\ref{detg}), we can write
    $$
{\cal Q} (f_n) \psi (y)\: \lambda_0\otimes\mu_0=
f_{(n)}^{\alpha_1 \alpha_2 \ldots \alpha_n} \: \frac{\partial^n
(|g|^{1/4}\psi (y))}{\partial{\alpha_1}\partial{\alpha_2} \ldots
\partial{\alpha_n}}\: \lambda_0\otimes\mu_0.
    $$
Now, in order to obtain the final result one have simply to rewrite
the last formula in a covariant way as it was shown in previous
section.


\section*{Acknowledgments}

The author is indebted to A. Aminova, I. Mikytiuk, and E. Patrin for
comments, suggestions and useful discussions. The work was partially
supported by Russian Foundation for Fundamental Investigation.


\end{document}